\documentclass[a4paper,twocolumn,11pt,accepted=2024-04-18]{quantumarticle}
\pdfoutput=1
\usepackage[utf8]{inputenc}
\usepackage[english]{babel}
\usepackage[T1]{fontenc}
\usepackage{amsmath}
\usepackage{hyperref}
\usepackage[numbers,sort&compress]{natbib}

\usepackage{tikz}
\usepackage{lipsum}

\usepackage{physics}
\usepackage{bm}
\usepackage[T1]{fontenc}
\usepackage{amsmath,amsfonts,bbm, color}
\usepackage{graphicx}
\usepackage{epstopdf}
\usepackage{amssymb}
\usepackage{appendix}
\usepackage{mathtools}
\usepackage{dsfont}
\usepackage{float}
\usepackage{relsize}
\usepackage{xcolor}
\usepackage{scalerel}
\DeclareMathOperator\artanh{artanh}

\usepackage{soul}

\begin{document}

\title{Quantum time dilation in a gravitational field}

\author{Jerzy Paczos}
\email{jerzy.paczos@fysik.su.se}
\affiliation{Department of Physics, Stockholm University, SE-106 91 Stockholm, Sweden}
\orcid{0000-0002-0674-9819}

\author{Kacper~D\k{e}bski}
\email{kdebski@fuw.edu.pl}
\affiliation{Institute of Theoretical Physics, University of Warsaw, Pasteura 5, 02-093 Warsaw, Poland}
\orcid{0000-0002-8865-9066}

\author{Piotr~T.~Grochowski}
\email{piotr.grochowski@uibk.ac.at}
\affiliation{Center for Theoretical Physics, Polish Academy of Sciences, Aleja Lotnik\'ow 32/46, 02-668 Warsaw, Poland}
 \affiliation{Institute for Quantum Optics and Quantum Information of the Austrian Academy of Sciences, A-6020 Innsbruck, Austria}
 \affiliation{Institute for Theoretical Physics, University of Innsbruck, A-6020 Innsbruck, Austria}
\orcid{0000-0002-9654-4824} 
 
\author{Alexander~R.~H.~Smith}
\email{arhsmith@anselm.edu }
\affiliation{Department of Physics, Saint Anselm College, Manchester, New Hampshire 03102, USA} 
\affiliation{Department of Physics and Astronomy, Dartmouth College, Hanover, New Hampshire 03755, USA} 
\orcid{0000-0002-4618-4832}
 
\author{Andrzej~Dragan}
\email{dragan@fuw.edu.pl}
\affiliation{Institute of Theoretical Physics, University of Warsaw, Pasteura 5, 02-093 Warsaw, Poland}
\affiliation{Centre for Quantum Technologies, National University of Singapore, 3 Science Drive 2, 117543 Singapore, Singapore}
\orcid{0000-0002-5254-710X}

\maketitle

\begin{abstract}
According to relativity, the reading of an ideal clock is interpreted as the elapsed proper time along its classical trajectory through spacetime.
In contrast, quantum theory allows the association of many simultaneous trajectories with a single quantum clock, each weighted appropriately.
Here, we investigate how the superposition principle affects the gravitational time dilation observed by a simple clock\,---\,a decaying two-level atom.
Placing such an atom in a superposition of positions enables us to analyze a quantum contribution to a classical time dilation manifest in spontaneous emission.
In particular, we show that the emission rate of an atom prepared in a coherent superposition of separated wave packets in a gravitational field is different from the emission rate of an atom in a classical mixture of these packets, which gives rise to a quantum gravitational time dilation effect. We demonstrate that this nonclassical effect also manifests in a fractional frequency shift of the internal energy of the atom that is within the resolution of current atomic clocks. In addition, we show the effect of spatial coherence on the atom's emission spectrum.
\end{abstract}

\section{Introduction}

The key insight of relativity is that time is measured by physical systems serving as clocks, and the relative flow of time between clocks depends on their velocity and distance from massive bodies.
When combined with quantum theory, it is natural to ask: What time does a clock measure if it moves in a superposition of velocities or is placed in a superposition of locations experiencing two different gravitational fields? 
Heuristically, such a clock would be expected to experience a superposition of proper times.

It is important to investigate the empirical consequences of such proper time superpositions because any associated phenomena are a consequence of both quantum theory and relativity, and thus, there is hope for a new test of physics at their intersection.
While attempts to detect phenomena originating from the combination of quantum mechanics and gravity have been made for many years (e.g., ~\cite{Colella, Peters, Muller}), examining interference phenomena for observing an effect of proper time superposition was developed only recently~\cite{Vedral2008,Zych2011,Bushev2016,Loriani2019,Roura2020}.
A common approach is based on an atomic interferometer in which the elapsed proper time along each arm of the interferometer is different due to gravitational time dilation.
If the atoms in the interferometer carry an internal clock degree of freedom, which-path information will be encoded in the clock degree of freedom, resulting in a decrease in interference visibility as a consequence of each arm experiencing different proper times.
A related special relativistic time dilation interferometer experiment has also been proposed~\cite{Bushev2016}.
In addition, quantum versions of the twin paradox~\cite{Vedral2008,Lindkvist2014,Lorek2015} and sequential boosts of quantum clocks~\cite{Paige2020} have also been shown to lead to nonclassical effects.
However, the experimentally realized atom interferometers have not operated with precise enough transitions to function as internal clocks capable of measuring proper time.
In contrast, we propose an experimentally viable setup in which the notion of proper time superposition is made operationally precise and is within current state-of-the-art experimental capabilities in atomic and ionic clocks.

A probabilistic notion of time dilation between quantum clocks was developed, which was used to examine relativistic time dilation between clocks moving in superpositions of momenta~\cite{Smith2020}. By modeling clocks as massive particles with internal clock degrees of freedom on which covariant time observables could be defined~\cite{Holevo2011,Busch1995}, it was shown that a clock moving in a superposition of two momentum wave packets relative to a laboratory frame that measures the time $t$ will measure an average proper time of
\begin{equation}\label{time_dilation}
    \langle{T}_\text{clock}\rangle=\left( \gamma_\text{C}^{-1}+ \gamma_\text{Q}^{-1}\right)t,
\end{equation}
where $  \gamma_\text{C}^{-1}$ corresponds to a statistical mixture of the classical time dilation expected by a classical clock moving with the average momentum of each wave packet, and $ \gamma_\text{Q}^{-1}$ is a consequence of the quantum coherence among the superposed wave packets. Equation~\eqref{time_dilation} represents the quantum analog of the special relativistic time dilation formula relating the time read by two clocks in relative motion.
It can be seen that the term $ \gamma_\text{Q}^{-1}$ represents a quantum correction to the average time dilation that depends on the relative phase of the wave packets, leading to so-called quantum time dilation.
Such superpositions of clocks have been interpreted within the context of quantum reference frame transformations and examined within a scalar field model~\cite{Giacomini2022}.

\begin{figure}[t!]
    \centering
    \includegraphics[width=0.45\textwidth]{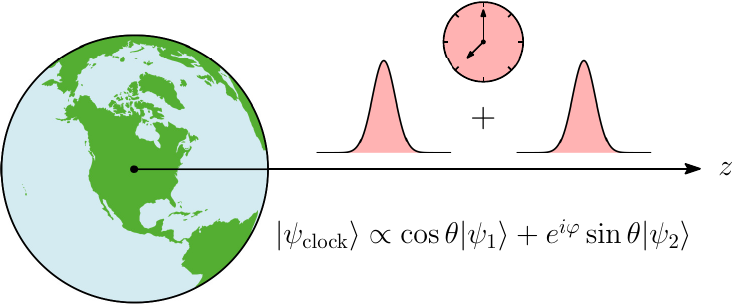}
    \caption{A pictorial representation of the considered setup in which a clock is placed in a superposition of different heights in a gravitational field as in Eq.~\eqref{clockSup1}.}
    \label{fig:setup}
\end{figure}

It was later shown that the same quantum time dilation effect was present in a more realistic clock model based on the spontaneous emission rate of an excited atom coupled to the electromagnetic field~\cite{Grochowski2021}, leading to a frequency shift on the order of up to $\delta \nu / \nu_0 \sim 10^{-15}$.
Modifications to the atomic spectrum were also shown to manifest due to quantum coherence among the momentum wave packets.
It is natural to expect analogous spectroscopic signatures when an atom is in a spatial superposition in a gravitational field, as depicted in Fig.~\ref{fig:setup}, given that it would experience a superposition of gravitational time dilation.

Indeed, an analogous nonclassical gravitational time dilation effect was shown to manifest for a quantum clock in a spatial superposition at different heights in a gravitational field~\cite{Khandelwal2020}, and interpreted within the formalism of quantum reference frames~\cite{Giacomini2021}.
In particular, the average time read by such a clock is analogous to Eq.~\eqref{time_dilation}, but with $\gamma_\text{Q}^{-1}$ corresponding to quantum corrections due to gravity.
As in Eq.~\eqref{time_dilation} there are two terms contributing to the average time observed by the quantum clock: the first term is a classical contribution corresponding to a statistical mixture of the time dilation observed by clocks located at the mean position of each wave packet comprising the superposition.
The second term is a correction due to the quantum nature of the clock's spatial degrees of freedom.
Such quantum corrections have been investigated in various models (scalar field theory~\cite{Stritzelberger2020,Lopp2021} and full quantum electromagnetism~\cite{Rzazewski1992,Grochowski2021}) taking into account a variety of contributions to the clock's Hamiltonian (i.e., nonrelativistic kinetic energy~\cite{Rzazewski1992,Stritzelberger2020,Smith2020,Khandelwal2020,Lopp2021,Grochowski2021}, special relativistic corrections~\cite{Stritzelberger2020,Smith2020,Khandelwal2020,Lopp2021,Grochowski2021}, and general relativistic corrections~\cite{Khandelwal2020}), and have been shown to manifest in physically relevant observables (total transition rates~\cite{Rzazewski1992,Stritzelberger2020,Lopp2021,Grochowski2021}, emission line shapes~\cite{Rzazewski1992,Grochowski2021}, well-defined time observables~\cite{Smith2020,Khandelwal2020}).

Depending on the situation considered, different phenomena may follow, such as homogeneous broadening of emission lines if only kinetic energy is involved or (special relativistic) quantum time dilation when only special relativistic corrections are considered.
An analogous gravitational quantum time dilation should manifest for the spatial superposition of wave packets in a gravitational field.
Reference~\cite{Khandelwal2020} considers such a superposition, however, with additional contributions to the clock's Hamiltonian\,---\,nonrelativistic kinetic energy and a special relativistic correction\,---\,therefore providing quantum corrections in addition to the expected gravitational quantum time dilation effect.
However, one can still extract such a contribution (see Appendix \ref{appendix_comparison}), yielding
    \begin{equation}
        \gamma^{-1}_\text{Q} = \frac{g}{4c^2} \frac{\left(z_2-z_1\right) \cos\varphi\sin 4 \theta }{\cos\varphi \sin 2\theta+\mathrm{e}^{\frac{(z_2-z_1)^2}{4\Delta^2}}}.
    \end{equation}
Here $g$ is the local magnitude of the gravitational field at the height $z=0$, and the clock's external degrees of freedom are prepared in the state
    \begin{equation}
        \ket{\psi_{\rm clock}} \propto \cos \theta \ket{\psi_1} + e^{i\varphi} \sin \theta \ket{\psi_2},
        \label{clockSup1}
    \end{equation}
    where $\ket{\psi_i}$ denotes a Gaussian wave packet of width $\Delta$ localized around $z_i$.

The purpose of this article is to demonstrate that the same correction to the time observed by a clock in a spatial superposition in a gravitational field $\gamma^{-1}_\text{Q}$ manifests in the spontaneous emission rate of an excited atom, and that spatial coherence in the presence of a gravitational field affects the atom's emission spectrum.

We focus on the problem of spontaneous emission as a platform for the proof-of-principle explanation of the quantum gravitational time dilation effect.
It serves as a paradigmatic example of a quantum clock and provides a clear setup to analyze.
However, the quantum gravitational time dilation would also manifest universally in different setups, involving, e.g., ions in Paul traps or many-particle systems in optical lattices.
Indeed, it will be shown that $\gamma^{-1}_\text{Q}$ results in a fractional frequency shift, which is the measurable quantity of interest for ion clocks.

We begin by analyzing the spontaneous emission rate of a two-level atom whose center-of-mass degrees of freedom are quantized and placed into a spatial superposition in a gravitational field. We show that the spontaneous emission rate experiences the nonclassical gravitational effect described in Ref.~\cite{Khandelwal2020}. To do so, we quantize the electromagnetic field in an accelerated frame and make use of the equivalence of uniformly accelerating observers and observers at rest in a gravitational field.
The use of an accelerated reference frame serves for computational convenience~\cite{Maybee2019}.
Obtaining results consistent with \cite{Khandelwal2020}, we show that the nonclassical effect characterized by $\gamma^{-1}_\text{Q}$ is consistent with the equivalence between clocks in a constant gravitational field and those in uniform acceleration.
Finally, we estimate the magnitude of $\gamma^{-1}_\text{Q}$, concluding that this quantum time dilation effect may be observable with present-day technology
and illustrate how spatial coherence in a gravitational field leads to nonclassical signatures in the atom's emission spectrum.

\section{Model}\label{sec::model}

Analysis of composite quantum systems in gravitational fields is of fundamental theoretical interest as experiments are beginning to probe regimes in which a description based on adding a background Newtonian gravitational potential to the Schr\"odinger equation is not enough~\cite{Colella,Kasevich1992}.
Specifically, coupling between internal and center-of-mass degrees of freedom has no classical analogue and implies, e.g., gravitationally induced quantum dephasing~\cite{Zych2011,Pikovski2015,Pang2016}, interferometric gravitational wave detection~\cite{Gao2018}, quantum tests of the classical equivalence principle~\cite{Schlippert2014}, and proposals for quantum versions of the equivalence principle~\cite{Viola1997,Anastopoulos2018,Zych2018}.

Usually, accounting for relativistic corrections in light-matter interactions is achieved through the addition of effects known from classical physics, such as second-order Doppler shifts, velocity-dependent masses, and time dilation due to relative motion or gravitational effects.
Such approaches are dangerous on their own because they do not guarantee self-consistency and usually rely on classical concepts such as worldlines or redshifts.
One example of problems arising in such a context includes spurious `friction' experienced by a moving and decaying atom~\cite{Sonnleitner2017}, which was later resolved by a proper relativistic treatment~\cite{Sonnleitner2018}.
For the gravity-free case, atom-field interactions have been rigorously derived to leading relativistic order~\cite{Sonnleitner2018}, while the extension to gravity was initially done by Marzlin in 1995~\cite{Marzlin1995}; however, the full first-order post-Newtonian expansion has only recently been presented by Schwartz and Giulini~\cite{Schwartz2019}. This derivation was further extended to the systems of nonzero total charge and used to obtain relativistic frequency shifts of ionic clocks in a rigorous way \cite{Martinez-Lahuerta2022}.

Such approaches assume that the atom is treated as a composite system and that the energy scales involved are below the threshold for pair production for any of the massive particles involved.
As such, this allows for a simplified relativistic analysis by performing quantization after putting the classical system in a fixed particle sector. In the case of a simple atomic model involving two charged, massive, and moving particles interacting with themselves via Coulomb forces and embedded in electromagnetic and weak gravitational fields, the effective atomic Hamiltonian was derived in~\cite{Schwartz2019, Zych2019}.
Under the assumption that the atom is heavy (or equivalently that effects due to the gravitational field dominate over the velocity spread of the atom), the Hamiltonian simplifies to the expected form:
\begin{align}\label{pn-hamiltonian-1}
\hat{H}&=M c^{2}\left(1+\frac{\phi(\hat{\boldsymbol{R}})}{c^{2}}\right)+\hbar \Omega\left(1+\frac{\phi(\hat{\boldsymbol{R}})}{c^{2}}\right)|e\rangle\langle e| \nonumber \\ &\quad +\hat{H}_{\text{L}}-\hat{\boldsymbol{d}} \cdot \hat{\boldsymbol{E}},
\end{align}
where $M$ is the mass of the atom, $\phi(\hat{\boldsymbol{R}})$ is the scalar gravitational potential in a post-Newtonian expansion, $\hat{\boldsymbol{R}}$ is the center-of-mass position of the atom, $\hbar\Omega$ is the energy gap of the relevant transition in the atom, $\ket{e}$ corresponds to the excited state of the atom, $\hat{H}_{\text{L}}$ is the Hamiltonian of the electromagnetic light in the presence of the gravitational field, and the last term describes the dipole coupling between the atom and electric field $\hat{\boldsymbol{E}}$.
The dipole interaction term retains its standard form if all the quantities involved (dipole moment $\hat{\boldsymbol{d}}$ and electric field $\hat{\boldsymbol{E}}$) are expressed as measured quantities and with respect to the proper time of an observer at $\boldsymbol{R}$~\cite{Marzlin1995,Lammerzahl1995,Schwartz2019}. The second term in Eq.~\eqref{pn-hamiltonian-1} can be interpreted as the Hamiltonian of the mass defect, which describes the relativistic coupling of the center of mass and the internal degrees of freedom \cite{Martinez-Lahuerta2022,Zych2011,Pikovski2015,Zych2018,Zych2019}. Therefore, it can be absorbed into the Hamiltonian of the center of mass (the first term in Eq.~\eqref{pn-hamiltonian-1}), with the mass effectively corrected by the mass defect. In the following analysis, we will ignore the binding potentials necessary to localize the wave packets; however, such potentials can be consistently taken into account within this model~\cite{Martinez-Lahuerta2022} and will not affect the signature of quantum time dilation, $\gamma_Q^{-1}$.

In the case of a homogeneous gravitational field, the system can be described in a uniformly accelerated (Rindler) frame.
Such a geometry allows us to describe a spectroscopic measurement that is performed close to the Earth's surface with the atomic cloud coherently delocalized in a gravitational field. We will re-express the total Hamiltonian in these coordinates to simplify the calculations and interpretation of the results. We start our analysis with a brief review of Rindler coordinates and provide a heuristic derivation of the total Hamiltonian in this frame.

\subsection{Rindler coordinates}\label{sec::rindler}

An accelerating frame of reference can be expressed via the following coordinate transformation \cite{Dragan2021}
\begin{equation}\label{rindler1}
    ct=\chi\sinh\left(\frac{g\tau}{c}\right),\quad z=\chi\cosh\left(\frac{g\tau}{c}\right),
\end{equation}
where $(ct, z)$ are Minkowski coordinates, $\tau$ is the Rindler time, $\chi$ is the Rindler distance, and $g$ is a reference proper acceleration corresponding to an observer measuring proper time $\tau$.
For simplicity, we restrict our considerations to only positive Rindler distance.
Then, observers characterized by fixed Rindler coordinates $\chi$ have constant proper acceleration $c^2/\chi$, and their proper times can be related to the parameter $g$ through $\frac{g \chi}{c^2} \tau$. 
They all share the common causally inaccessible region lying beyond the hypersurface $z=ct$, known as the Rindler horizon.
The inverse transformation reads 
\begin{equation}\label{rindler2}
    c\tau=\frac{c^2}{g}\artanh\left(\frac{ct}{z}\right),\quad \chi=\sqrt{z^2-c^2t^2}.
\end{equation}
It is also customary to use the so-called radar coordinates $(c\tau,\xi)$ with $\xi$ defined by $\chi=\frac{c^2}{g}\,\mathrm{e}^{\,g\xi/c}$.
In particular, these were used in \cite{Maybee2019} to perform the quantization of the electromagnetic field in a uniformly accelerated reference frame.

\subsection{The Hamiltonian}\label{sec::hamiltonian}

We now proceed to construct the Hamiltonian describing a two-level atom interacting with the electromagnetic field in Rindler coordinates that mimics the effect of a homogeneous gravitational field.
First, in the absence of gravity, the Hamiltonian of an atom of mass $M$, with ground state $\ket{g}$ and excited state $\ket{e}$ separated by an energy difference $\hbar\Omega$, is given by
\begin{equation}
    \hat{H}^{(0)}=\hat{H}_{\text{atom}}^{(0)}+\hat{H}_{\text{field}}^{(0)}+\hat{H}_{\text{af}}^{(0)},
\end{equation}
with the terms correspond to the atomic Hamiltonian, electromagnetic field Hamiltonian, and atom-field coupling, respectively, all specified in the Jaynes-Cummings model \cite{Gerry2004, Lambropoulos2007, Grynberg2010, Rzazewski1992}
\begin{equation}
\begin{split}
    \hat{H}_{\text{atom}}^{(0)}&=Mc^2+\hbar\Omega\ket{e}\!\bra{e},\\
    \hat{H}_{\text{field}}^{(0)}&=\sum_{\boldsymbol{k},\lambda}\hbar\omega_{\boldsymbol{k}}\hat{a}^\dagger_{\boldsymbol{k},\lambda}\hat{a}_{\boldsymbol{k},\lambda},\\
    \hat{H}_{\text{af}}^{(0)}&=-\hat{\boldsymbol{d}}\cdot\hat{\boldsymbol{E}}.
    \label{eqn:H_atom(0)}
\end{split}
\end{equation}
Here $\hat{a}_{\boldsymbol{k},\lambda}^\dagger$ and $\hat{a}_{\boldsymbol{k},\lambda}$ are the creation and annihilation operators, respectively, of a photon with wave vector $\boldsymbol{k}$  (eigenfrequency $\omega_{\boldsymbol{k}}$) and polarization $\lambda$, $\hat{\boldsymbol{d}}=\boldsymbol{d}(\ket{e}\!\bra{g}+\ket{g}\!\bra{e})$ is the dipole operator of the atom, $\hat{\boldsymbol{E}}$ is the electric field operator, and the atomic Hamiltonian has been shifted by the atom's rest mass energy.

We now find the corresponding terms expressed in Rindler coordinates. At first, we consider the rest energy of the atom (the $\hat{H}_{\text{atom}}^{(0)}$ term). Because we are employing Rindler coordinates, the metric tensor $g_{\mu\nu}$ is diagonal and only the $g_{00}$ component is nontrivial. Let us consider coordinates $\left(x^0,x^i\right)$ and take the metric tensor to be $g_{\mu\nu}=g_{00}\delta_{\mu 0}\delta_{\nu 0}-\delta_{\mu i}\delta_{\nu j}\delta^{ij}$. Classically, the Lagrangian of a free particle can then be written as
\begin{equation}
    \mathcal{L}=-H_{\text{atom}}^{(0)}\sqrt{g_{\mu\nu}\Dot{x}^\mu\Dot{x}^\nu}=-H_{\text{atom}}^{(0)}\sqrt{g_{00}-\Dot{\bar{x}}^2},
\end{equation}
where $H_{\text{atom}}^{(0)}$ is the classical counterpart of $\hat{H}_{\text{atom}}^{(0)}$, $\Dot{x}^\mu=\frac{dx^\mu}{dx^0}$, and $\Dot{\bar{x}}^2=\delta_{ij}\frac{dx^i}{dx^0}\frac{dx^j}{dx^0}$. Here, we should stress that $\Dot{\bar{x}}$ is not the velocity of the particle but the velocity divided by $c$ because the coordinate $x^{0}$ has a unit of length. For a static metric, i.e., if $g_{00}$ does not depend on $x^0$ (e.g., the Rindler metric), the following quantity (Hamiltonian) is conserved~\cite{Zych2019}
\begin{align}
H_{\text{atom}}&=\frac{\partial \mathcal{L}}{\partial \Dot{x}^{i}} \Dot{x}^{i}-\mathcal{L}
=\frac{H_{\text{atom}}^{(0)} g_{00}}{\sqrt{g_{00}-\Dot{\bar{x}}^{2}}}.
\end{align}
Here we treat $\Dot{x}^i$'s as functions of the canonical momenta $p_i=\frac{\partial \mathcal{L}}{\partial \Dot{x}^{i}}$.
Finally, if the particle is at rest, its energy is given by $H_{\text{atom}}=H_{\text{atom}}^{(0)}\sqrt{g_{00}}$. We obtain a quantum version of this equality by simply replacing the classical Hamiltonians on both sides with their corresponding operators, i.e., $\hat{H}_{\text{atom}}=\hat{H}_{\text{atom}}^{(0)}\sqrt{g_{00}}$. In further considerations, we restrict our calculations to the stationary case. 

We now return to the analysis of the Rindler metric for which $g_{00}=\left(\frac{g\chi}{c^2}\right)^2$. We introduce the parameter $z$ as a distance between the particle and the reference hyperbola given by $\chi=\frac{c^2}{g}+z$. One can rewrite the Hamiltonian as
\begin{equation}\label{atomic_hamiltonian}
    \hat{H}_{\text{atom}}=\hat{H}_{\text{atom}}^{(0)}\frac{g\chi}{c^2}=\hat{H}_{\text{atom}}^{(0)}\left(1+\frac{\phi(z)}{c^2}\right),
\end{equation}
where $\phi(z)=gz$ is the linear gravitational potential. Therefore, the atomic energy scales by a factor $1+\phi(z)/c^2$. In addition, it is worth noting that this result agrees with \cite{Martinez-Lahuerta2022}, which strongly justifies our simplified method based on treating gravity in a fully kinematic way by using a Rindler frame of reference. We emphasize that this scaling factor appears in our calculations naturally without any additional reasoning.
One sees that the inclusion of this factor couples the internal clock degrees of freedom with the motional degrees of the atom through the term, $\hat{H}_{\text{atom}}^{(0)} \phi(z)/c^2$, which is responsible for the gravitational time dilation felt by the clock~\cite{Zych2011,Pikovski2015,Zych2018,Zych2019}.

It should be emphasized that the atomic Hamiltonian should, in general, contain a kinetic term $\boldsymbol{P}^2/2M$, where $\boldsymbol{P}$ is the total momentum of the atom \cite{Smith2020,Pikovski2015}. However, we assume that the atom is very heavy, so the kinetic energy related to both the total momentum of the system and its dispersion is negligible compared to other kinds of energy described by the Hamiltonian. This means we will not consider any motion of the center of mass. Therefore, we will discard all the terms depending on either the velocity or its dispersion derived in~\cite{Khandelwal2020}. We can additionally justify this omission by estimating the magnitude of individual time dilation effects from~\cite{Khandelwal2020}.
We perform such estimation in Appendix~\ref{appendix_comparison} and show that, for the considered range of parameters, the corrections due to motion are indeed much smaller than the purely gravitational correction, thus justifying our simplifying assumptions.

The electromagnetic field Hamiltonian in curved spacetime was derived in \cite{Maybee2019}. Here we consider only one (right) Rindler wedge; hence, we neglect all terms containing ladder operators from the left Rindler wedge, and we use the following Hamiltonian
\begin{equation}\label{field}
    \hat{H}_{\text{field}}=\sum_{\lambda=1,2} \int_{-\infty}^{\infty}\mathrm{d}k\, \hbar\omega_{k}\hat{b}_{k,\lambda}^{\dagger R} \hat{b}_{k,\lambda}^{R},
\end{equation}
where $\hat{b}_{k,\lambda}^{\dagger R}$ and $\hat{b}_{k,\lambda}^{R}$ are, respectively, the raising and lowering operators in the right Rindler wedge, and $k=|\boldsymbol{k}|$ is the wavenumber.
We will assume that the atom is placed in an optical cavity, which allows photons to propagate only in the direction of the gravitational field.
Therefore, we reduce our problem to a single spatial dimension, making the quantization scheme easier~\cite{Maybee2019}.

To express the atom-field interaction term in Rindler coordinates we only need to express the electric field in Rindler coordinates. The absence of the $\sqrt{g_{00}}$ term in this interaction Hamiltonian is a consequence of the fact that we will use so-called coordinate components of the electric field~\cite{Schwartz2019}. For this purpose, we express the electric field as~\cite{Maybee2019}
\begin{equation}
\boldsymbol{\hat{E}}=i \sum_{\lambda=1,2} \int_{-\infty}^{\infty}\mathrm{d}k\, \sqrt{\frac{\hbar\omega_{k}}{4 \pi\varepsilon_0}}\left[\mathrm{e}^{ik\xi} \hat{b}_{k,\lambda}^{R}+\text { H.c. }\right] \hat{\boldsymbol{e}}_{\lambda}.
\end{equation}
Again, we have ignored the terms containing ladder operators from the left Rindler wedge. Notice that in contrast to \cite{Maybee2019}, we work in the Schr\"odinger picture, so the electric field operator has no explicit time dependence. Employing the rotating wave approximation, the interaction term simplifies to
\begin{equation}
\begin{split}
    \hat{H}_{\text{af}}&=-\boldsymbol{\hat{d}}\cdot\boldsymbol{\hat{E}}\\
    &=-i\hbar \!\sum_{\lambda=1,2}g_{k,\lambda} \int_{-\infty}^{\infty} \!\mathrm{d} k \left(\mathrm{e}^{i k\xi} \hat{b}_{k,\lambda}^{R}|e\rangle\langle g|
 \!+\! \text{H.c.}\right)\!,
\end{split}
\end{equation}
where
\begin{equation}
g_{k,\lambda}=\sqrt{\frac{\omega_{k}}{4 \pi\hbar\varepsilon_0}}\boldsymbol{d}\cdot\hat{\boldsymbol{e}}_{\lambda}
\end{equation}
is the coupling constant governing the strength of the atom-light interaction.

Comparing the final Hamiltonian in Rindler coordinates
\begin{equation}\label{rindlerham}
    \hat{H}=\hat{H}_{\text{atom}}+\hat{H}_{\text{field}}+\hat{H}_{\text{af}},
\end{equation}
with the simplified version of the Hamiltonian in Eq.~\eqref{pn-hamiltonian-1}, one concludes that they are the same. Therefore, we investigate spontaneous emission in a uniformly accelerating reference frame and expect that the associated results will agree with a post-Newtonian gravitational analysis.

\section{\label{sec:emission}Spontaneous emission in the gravitational field}
We consider a setting in which a two-level atom is at rest at some height above ground level $z=0$ ($\chi=\frac{c^2}{g}$) in a gravitational field. Using Rindler coordinates, we choose the reference hyperbola to be located at ground level, which means that the time coordinate $\tau$ is the proper time of an observer at $z=0$. We assume that the action takes place in only one (right) Rindler wedge, far from the Rindler horizon.

\begin{figure*}[t]
\centering
  \includegraphics[width=0.3 \textwidth]{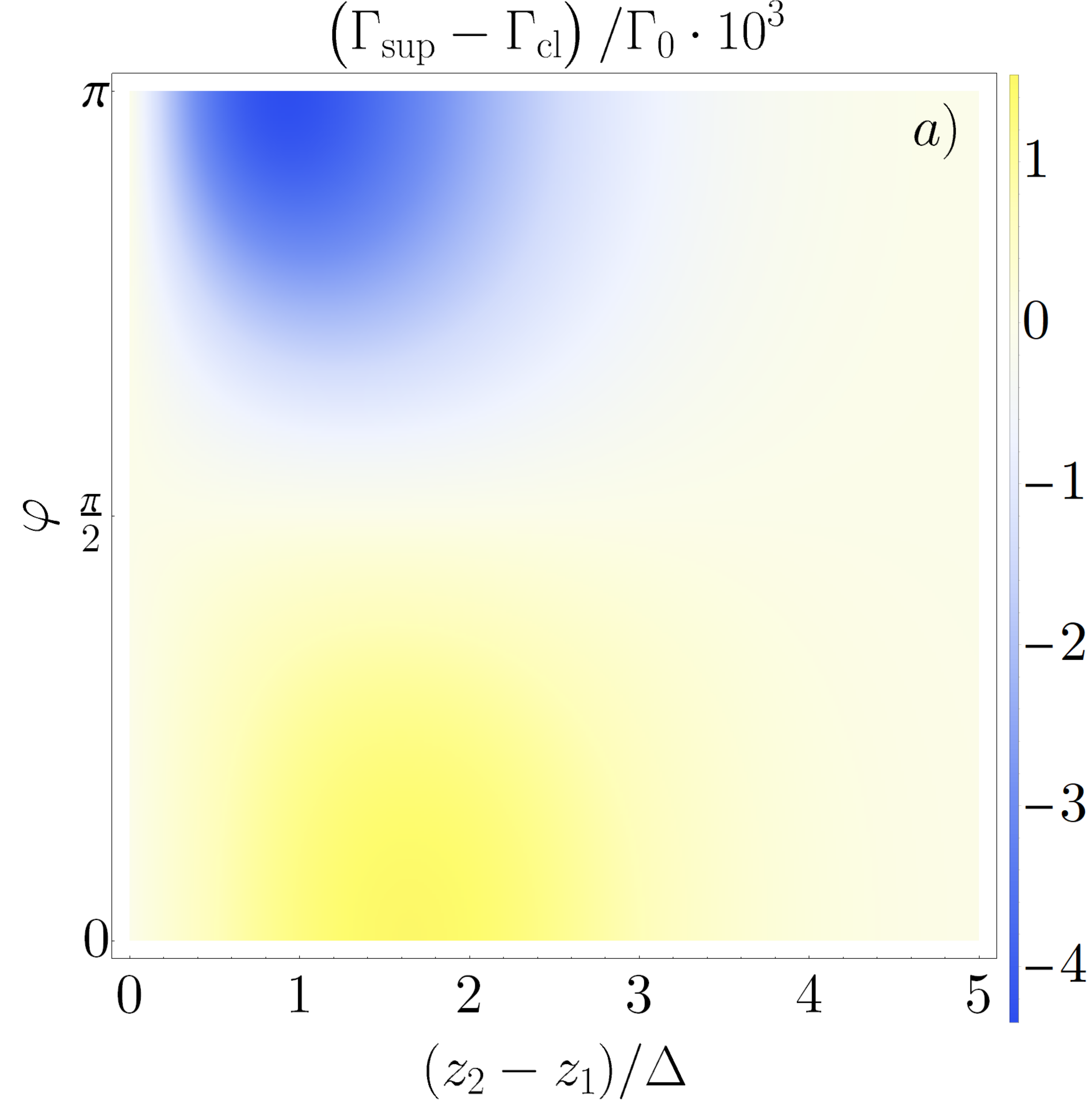} \quad \includegraphics[width=0.3\textwidth]{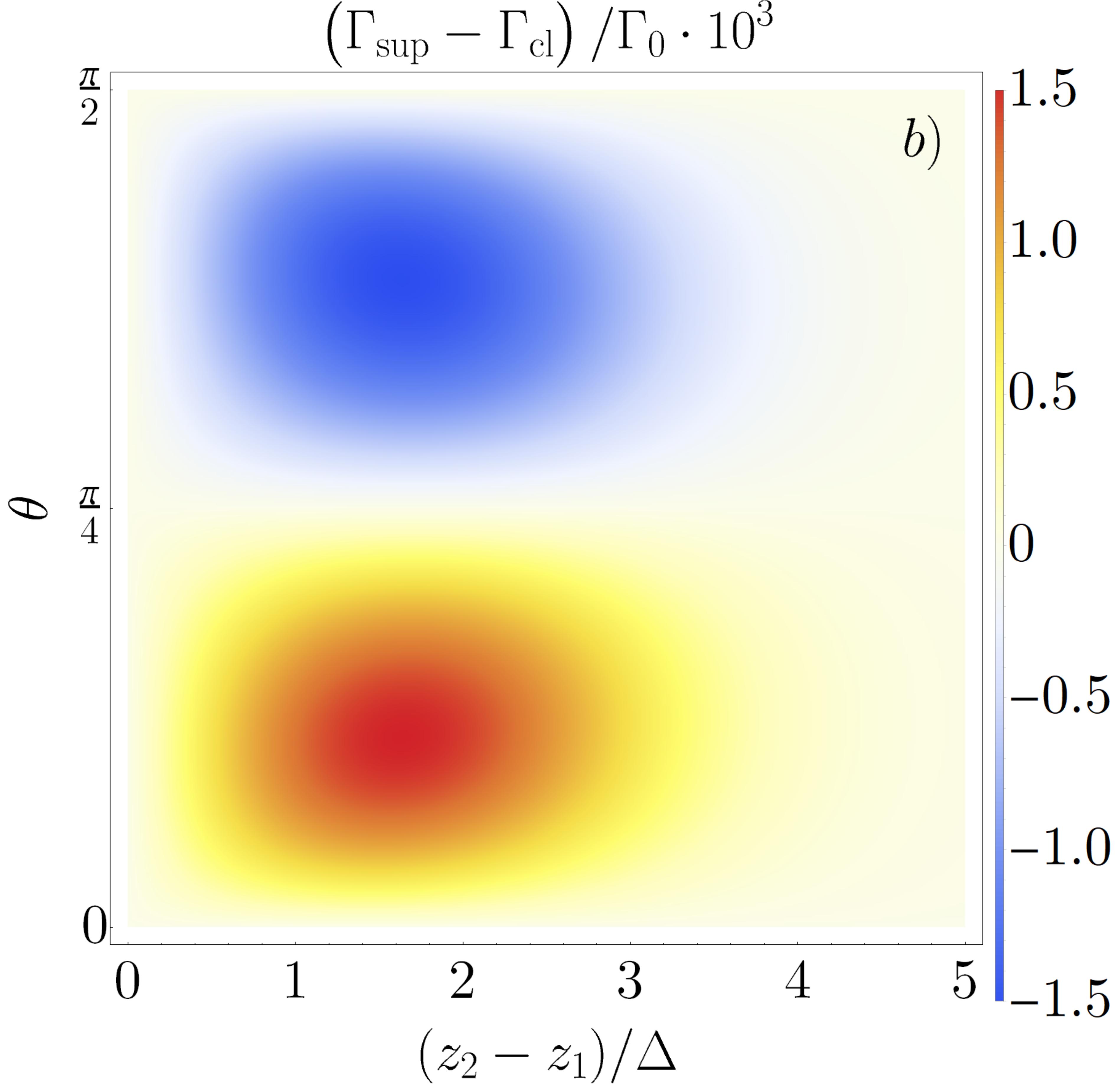} \quad \includegraphics[width=0.3\textwidth]{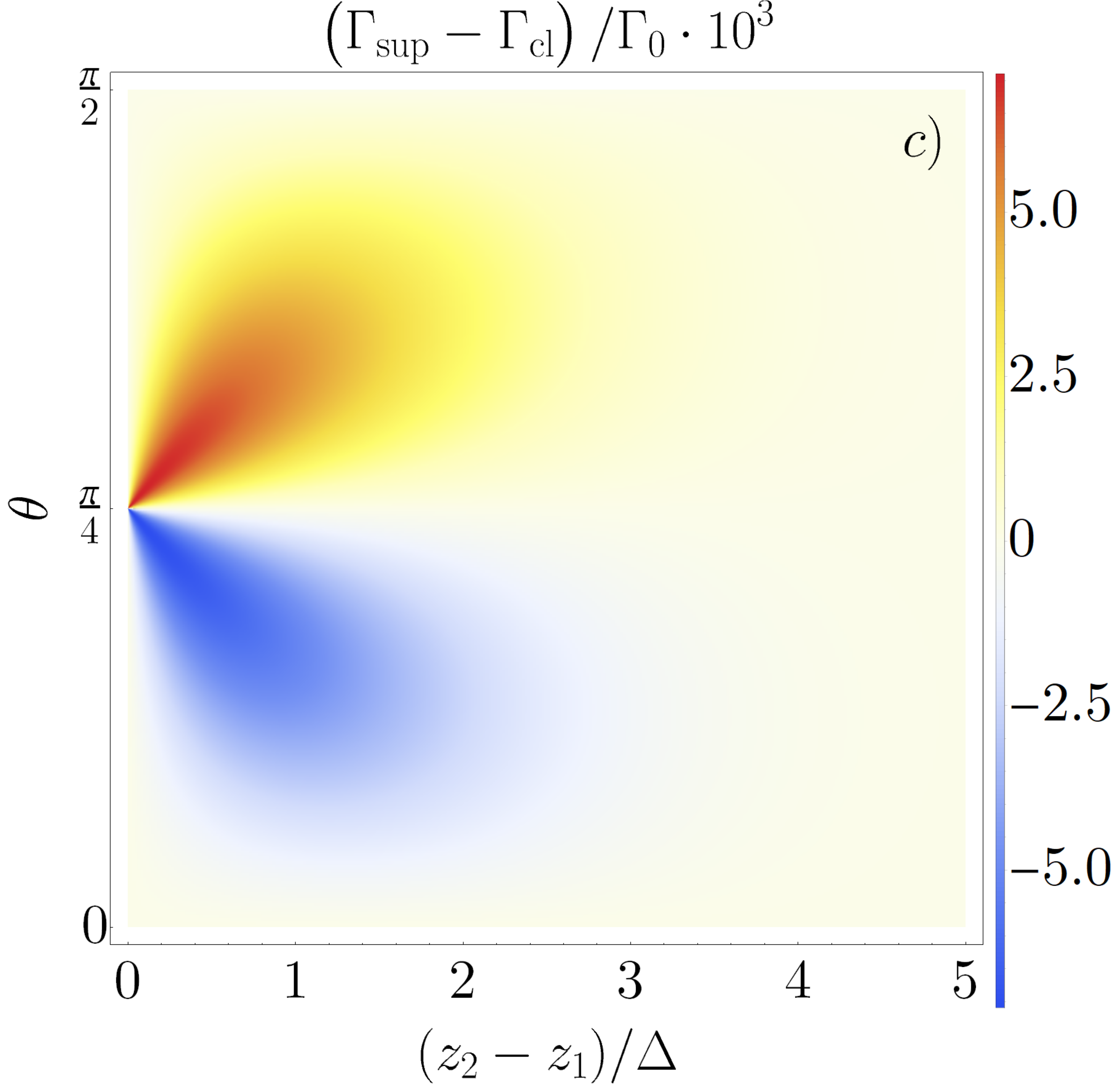}
\caption{The difference in total emission rate between a superposition and a classical mixture of two Gaussian wave packets localized around $z_1$ and $z_2$ as a function of the difference $z_2-z_1$, the relative phase $\varphi$, and weight $\theta$: (a) the wave packets have unequal weights ($\theta=\pi/8$), (b) relative phase is fixed at $\varphi=0$, (c) relative phase is fixed at $\varphi=\pi$. In each plot, the position spread along the $z$ axis is fixed $\Delta=0.01c^2/g$.}
\label{fig::transition_rate}
\end{figure*}

\begin{figure*}[t]
\centering
  \includegraphics[width=.45\textwidth]{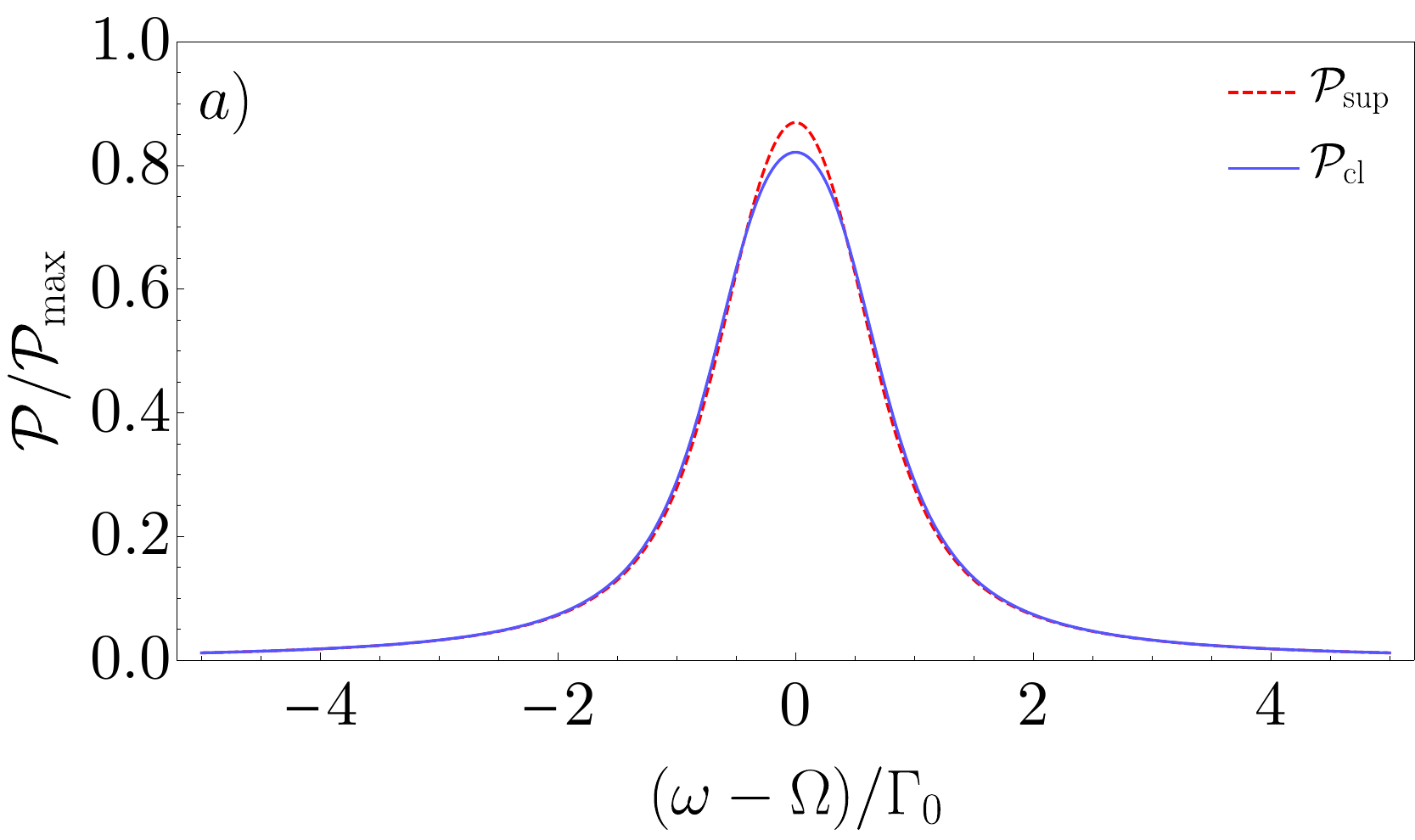} \qquad \includegraphics[width=.45\textwidth]{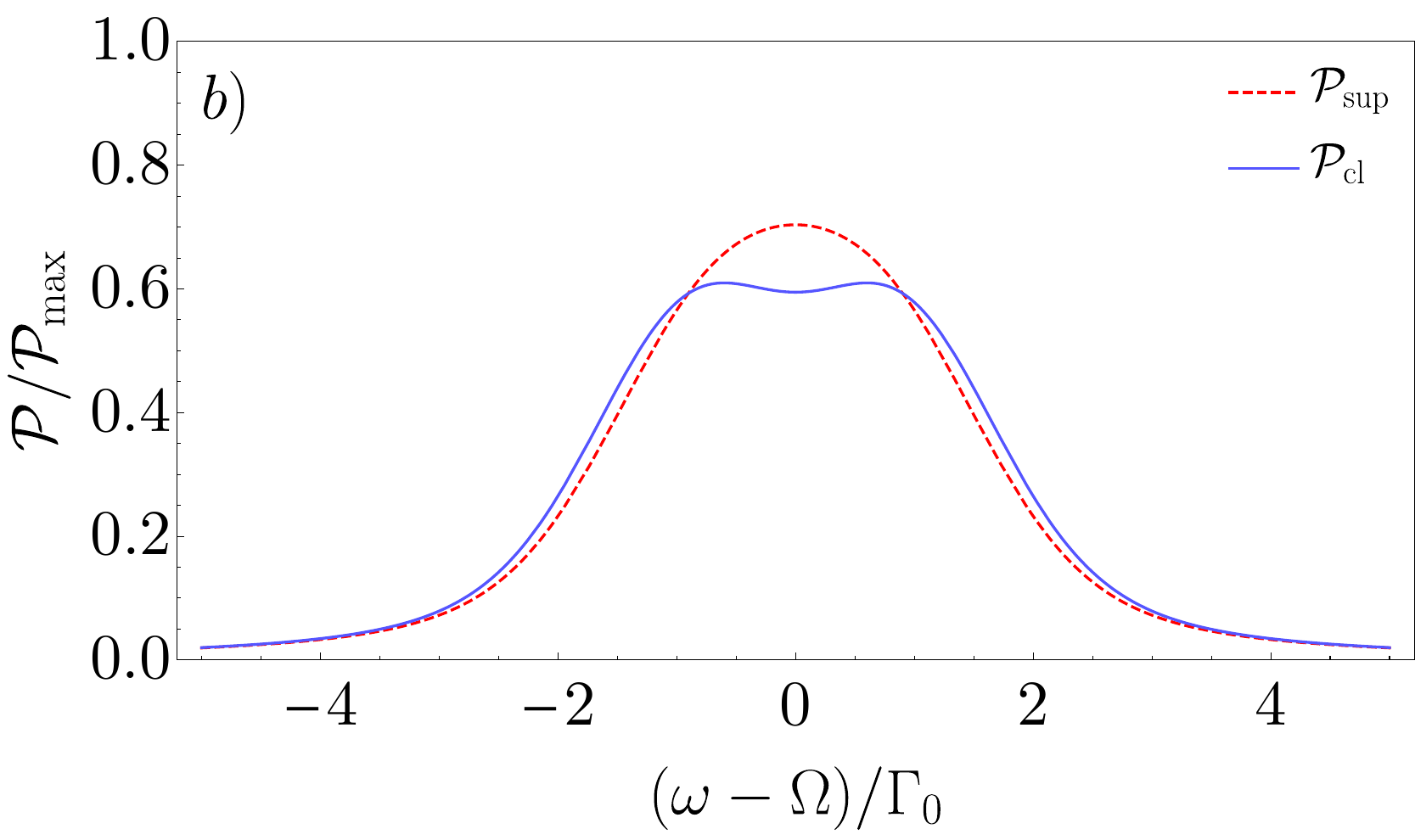}
 \includegraphics[width=.45\textwidth]{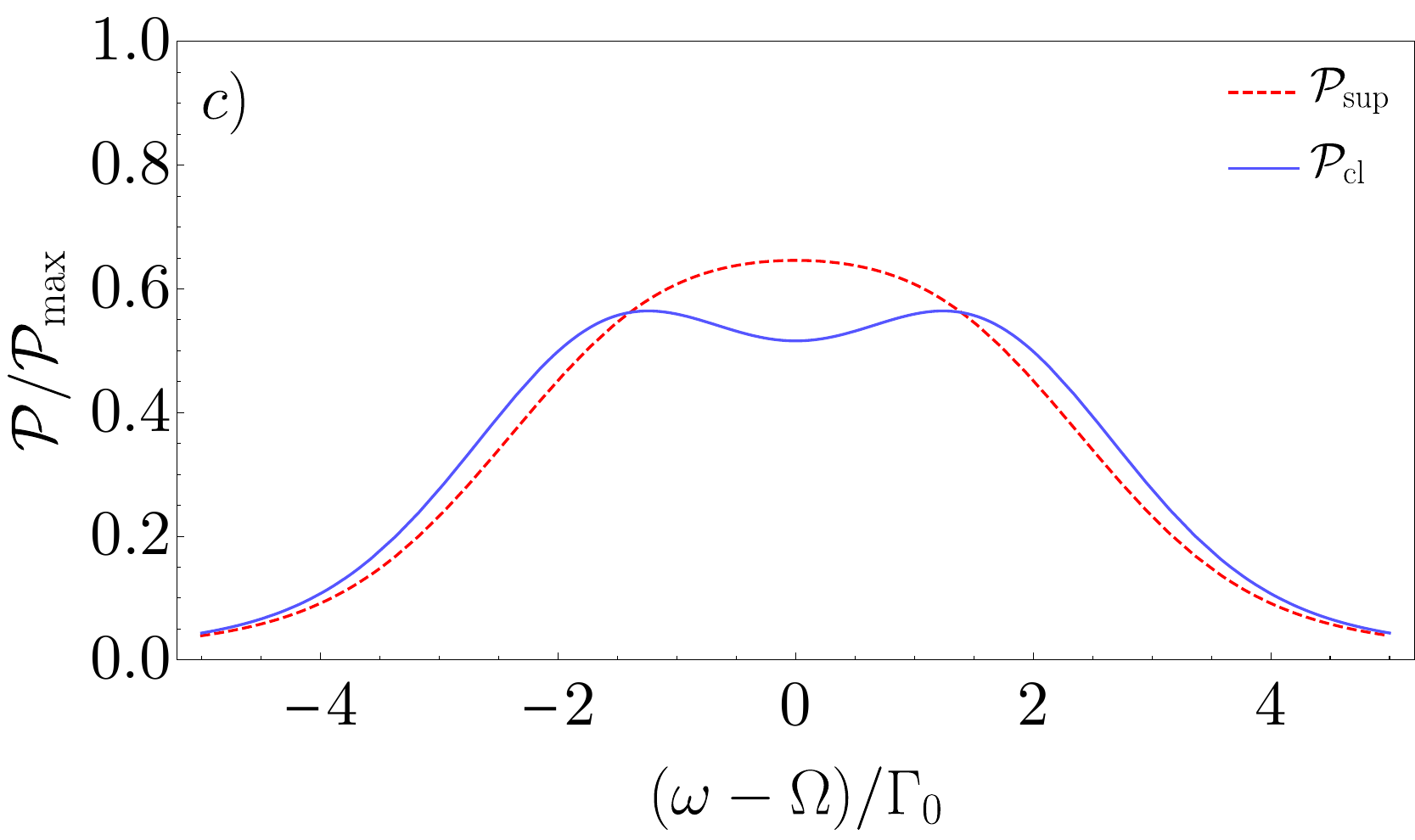} \qquad \includegraphics[width=.45\textwidth]{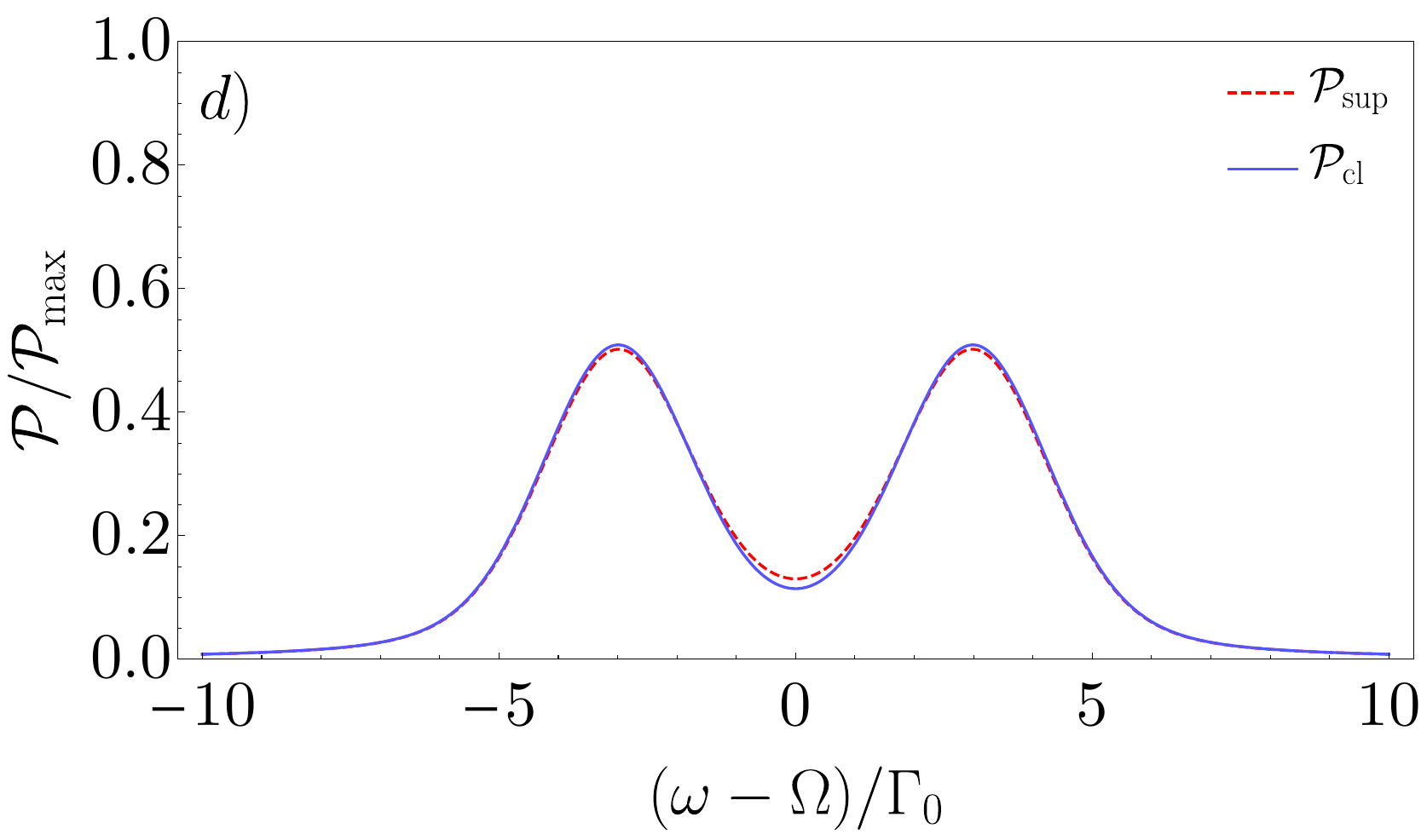}
\caption{Comparison of the emission line shape $\mathcal{P}(\omega)$ of an atom that is initially in a coherent superposition ($\mathcal{P}_\text{sup}$) and in a classical mixture ($\mathcal{P}_\text{cl}$) of wave packets peaked at different heights $z_1$ and $z_2$. The transition is assumed to be extremely narrow $\Omega/\Gamma_0\approx1.5\times10^{17}$ (e.g. aluminium ${}^1S_0-{}^1P_0$ transition). (a) is calculated for $z_1=-z_2=-2\times10^{-18}c^2/g$, (b) is associated with $z_1=-z_2=-6\times10^{-18}c^2/g$, while (c) corresponds to $z_1=-z_2=-10^{-17}c^2/g$. In these three cases, the spread of the wave packets equals $\Delta=(z_2-z_1)/2$. On the other hand, in case  (d), which is also plotted for $z_1=-z_2=-10^{-17}c^2/g$, the dispersion is equal to $\Delta=(z_2-z_1)/4$.}
\label{fig::spectral_lines}
\end{figure*}

We will describe the state of the system by a state vector providing information about the atom's position $z$, the internal state of the atom (ground $\ket{g}$ or excited $\ket{e}$), and the number of photons $n_{k,\lambda}$ in mode $k$ with polarization $\lambda$ in the considered Rindler wedge. Initially, the atom is excited and placed in an electromagnetic vacuum
\begin{equation}\label{initial_state}
\ket{\Psi(0)}=\int\mathrm{d}z\, \psi(z)\ket{z,e,0}.
\end{equation}
Because deexcitation of the atom can produce only one photon, the general state of the system at the Rindler time $\tau$ reads
\begin{equation}\label{general_state}
\begin{aligned}
\ket{\Psi(\tau)}=&\int \mathrm{d} z\, \alpha(z,\tau)\ket{z,e,0}\\
&+\sum_{\lambda=1,2}\int\int \mathrm{d} k\, \mathrm{d} z\, \beta_{k,\lambda}(z,\tau)\ket{z,g,1_{k,\lambda}},
\end{aligned}
\end{equation}
which has been expanded in the energy eigenstates $\ket{z,e,0}$ and $\ket{z,g,1_{k,\lambda}}$, associated respectively with energies
\begin{equation}
\begin{split}
\hbar\omega_g(z)&=Mc^2\left(1+\frac{gz}{c^2}\right),\\
\hbar\omega_e(z)&=\left(Mc^2+\hbar\Omega\right)\left(1+\frac{gz}{c^2}\right).
\end{split}
\end{equation}
The integrals over $z$ in Eqs.~\eqref{initial_state} and \eqref{general_state} should go from $z=-c^2/g$ (position $\chi=0$ corresponds to the event horizon) to $z=\infty$, but because of the assumption that the atom is far from the Rindler horizon, we can extend these integrals to $z=-\infty$. Notice that because we omitted the kinetic term $\boldsymbol{P}^2/2M$ in Eq.~\eqref{rindlerham}, it does not contain any momentum-dependent terms, and we can restrict our considerations to the position representation, which greatly simplifies the analysis.

We expect that, due to gravitational time dilation, the atom's lifetime will depend on the initial distribution of the state in the $z$ direction. This should be reflected in the dependence of the transition rate
\begin{equation}
    \Gamma=-\frac{\mathrm{d}}{\mathrm{d}\tau}\int \mathrm{d}z\,|\alpha(z,\tau)|^2,
\end{equation}
on the initial wave function $\psi(z)$.

The coefficients $\alpha(z,\tau)$, and $\beta_{k,\lambda}(z,\tau)$ appearing in Eq.~\eqref{general_state} are derived in Appendix~\ref{appendix::evolution}, with the result being:
\begin{align}
\alpha(z,\tau)&=\:\psi(z)\exp\left[\left(-i\omega_e(z) -\frac{\Gamma(z)}{2}\right)\tau\right],\label{alpha}\\ \nonumber
\beta_{k,\lambda}(z,\tau)&=\:\frac{g_{k,\lambda}\psi(z)\mathrm{e}^{-i\left(\omega_g(z)+k\xi\right)}}{\frac{1}{2}\Gamma(z)+i\left(\Omega\left(1+\frac{gz}{c^2}\right)-\omega_k\right)}\\
&\quad \times\left[\mathrm{e}^{-\left(i\Omega\left(1+\frac{gz}{c^2}\right)+\frac{1}{2}\Gamma(z)\right)\tau}-\mathrm{e}^{i\omega_k\tau}\right],\label{beta}
\end{align}
where $\Gamma(z)=\left(1+\frac{gz}{c^2}\right)\frac{\Omega \boldsymbol{d}^2}{2\hbar c\varepsilon_0}\equiv\left(1+\frac{gz}{c^2}\right)\Gamma_0$ is the transition rate of the atom localized at height $z$.
Then, the total transition rate for $\Gamma_0\tau\ll1$ reads (see Appendix~\ref{appendix::emission})
\begin{equation}\label{transition_rate}
\Gamma=\int \mathrm{d}z\,|\psi(z)|^2\Gamma(z),
\end{equation}
which is a weighted mean of transition rates at different heights.
Moreover, as shown in Appendix~\ref{appendix::emission}, the normalized standard deviation of the transition rate associated with the spatial extent is
\begin{equation}\label{standard deviation}
    \frac{\sigma_\Gamma}{\Gamma}=\frac{g\sigma_z}{c^2},
\end{equation}
where $\sigma_z$ is the standard deviation of the position space wave function $\psi(z)$.

Comparing the value of $\Gamma$ for specific initial states allows us to check whether there is any difference between the time dilation observed by atoms in coherent superpositions and in classical mixtures of position wave packets.
In the first case, the initial wave function is given by
\begin{equation}
\psi_{\text{sup}}(z)=\mathcal{N}\left[\cos\theta\mathrm{e}^{-\frac{(z-z_1)^2}{2\Delta^2}}+\mathrm{e}^{i\varphi}\sin\theta\mathrm{e}^{-\frac{(z-z_2)^2}{2\Delta^2}}\right]\!,
\end{equation}
where
\begin{equation}
\mathcal{N}=\left[\sqrt{\pi}\Delta\left(1+\cos\varphi\sin2\theta\mathrm{e}^{-(z_1-z_2)^2/4\Delta^2}\right)\right]^{-\frac{1}{2}}\!,
\end{equation}
whereas in the second case, the probability distribution reads
\begin{equation}
\mathnormal{P}_{\text{cl}}(z)=\frac{1}{\sqrt{\pi}\Delta}\!\left[\cos^2\theta\mathrm{e}^{-\frac{(z-z_1)^2}{\Delta^2}}\!+\sin^2\theta\mathrm{e}^{-\frac{(z-z_2)^2}{\Delta^2}}\right]\!.
\end{equation}
We evaluate Eq.~\eqref{transition_rate} and obtain the difference in the total emission rate between these two cases
\begin{align}\label{ratio}
\frac{\Gamma_{\text{sup}}-\Gamma_{\text{cl}}}{\Gamma_0}&=\!\int \mathrm{d}z\,\!\left(1+\frac{gz}{c^{2}}\right)\!\left[|\psi_{\text{sup}}(z)|^2-\mathnormal{P}_{\text{cl}}(z)\right] \nonumber\\
&= \frac{g}{4c^2} \frac{\cos\varphi\sin4\theta(z_2-z_1)}{\cos\varphi\sin2\theta+\mathrm{e}^{\frac{(z_2-z_1)^2}{4\Delta^2}}} \nonumber \\
&= \gamma^{-1}_\text{Q}.
\end{align}

Let us now argue that this result is universal.
As an example, Ref.~\cite{Martinez-Lahuerta2022} analyzed an ion clock setup from the perspective of fractional frequency shifts of internal energy levels, including relativistic and gravitational ones.
Following their reasoning (see Eq.~(37) in \cite{Martinez-Lahuerta2022}), the fractional frequency shift operator $\frac{\delta \hat{\nu}}{\nu_0}$ for a two-level system in a gravitational field consists of two terms: the center-of-mass kinetic energy and a contribution due to gravity.
With the approximation considered in our work\,---\,the rest mass energy of the atom much larger than its center-of-mass kinetic energy\,---\,this operator reads 
\begin{equation}
    \frac{\delta \hat{\nu}}{\nu_0} =  \frac{g \hat{z}}{c^2}.
\end{equation}
Then, the difference between expected fractional frequency shifts in the coherent and incoherent cases is given by 
\begin{equation}
 \expval{\frac{\delta \hat{\nu}}{\nu_0}}_{\text{sup}} -  \expval{\frac{\delta \hat{\nu}}{\nu_0}}_{\text{cl}} = \gamma^{-1}_\text{Q},
\end{equation}
which is exactly the contribution to the quantum gravitational time dilation effect.
One needs to note, however, that any specific system will involve a different analysis involving all the experimental frequency shifts, dynamics leading to different effective conditional states, etc.
Such analysis goes beyond the scope of this work, whose main aim is to show the proof-of-principle description of the quantum gravitational time dilation effect.

Let us discuss some basic properties of this result. First,  $\gamma^{-1}_\text{Q}$ vanishes when $\theta=0$, $\frac{\pi}{4}$, or $\frac{\pi}{2}$. If $\theta=0$ or $\theta=\frac{\pi}{2}$, there is no difference between the coherent superposition and the classical mixture; therefore, vanishing Eq.~\eqref{ratio} is something we should expect. Second, the quantity Eq.~\eqref{ratio} also vanishes for $\varphi=\frac{\pi}{2}$. This is the case when the probability distribution corresponding to the coherent superposition is the same as the probability distribution of the classical mixture. Finally, Eq.~\eqref{ratio} vanishes for $|z_2-z_1|\gg\Delta$, which corresponds to localized states whose separation from each other is large compared to their spread, so the interference effects are negligible. The correction Eq.~\eqref{ratio} is plotted in Fig.~\ref{fig::transition_rate} for fixed values of $\theta$ and $\varphi$. Note that $\gamma_{\rm Q}^{-1}$ can be either positive or negative, depending on the relative phase $\varphi$ and relative weight $\theta$.

The quantity Eq.~\eqref{ratio} seems to be extremely small because of the factor $\frac{g}{c^2}(z_2-z_1)$, which for the height difference of the wave packets $z_2-z_1\sim 1\text{cm}$ is of the order of $\sim 10^{-18}$. However, the same factor appears in classical gravitational time dilation\,---\,the difference of the time read by two clocks placed, respectively, at heights $z_2$ and $z_1$ is also proportional to $\frac{g}{c^2}(z_2-z_1)$. Therefore, provided that the spread of the wave packets is comparable to the distance between them, the effect of the quantum time dilation can be of the same order of magnitude as classical gravitational time dilation.

The spatial extent of a non-Gaussian wave function necessarily implies an additional source of uncertainty in the estimation of the transition rate (or fractional frequency shift) with respect to the normal scenario of a well-localized atom.
This correction to the total uncertainty can be approximated to leading order through Eq.~\eqref{standard deviation} and, in common experimental scenarios, is reduced through many experimental interrogations $N$, so that its contribution scales like $\sim \sigma_\Gamma / \sqrt{N}$.
In general, the total uncertainty depends on the specific experimental realization and involves many contributions, e.g., projection noise.
For example, a detailed analysis in~\cite{Martinez-Lahuerta2022} for trapped-ion optical clocks revealed that the spatial extent of the Gaussian wave packet is shown to have a subleading effect on the total precision.
In the case we consider here, the variance in the spatial probability distribution of a superposition of two Gaussian wave packets is
\begin{equation}
    \sigma_z^2 = \frac{1}{4} (z_2 - z_1)^2 + \frac{1}{2}\Delta^2 + \sigma_\text{Q}^2,
\end{equation}
where
\begin{equation}
    \sigma_\text{Q}^2 = -\frac{(z_2-z_1)^2}{4} \frac{\cos{ \varphi} \sin{2 \theta}}{e^{\frac{(z_2-z_1)^2}{4 \Delta^2}} + \cos{ \varphi} \sin{2 \theta} }
\end{equation}
is an additional quantum correction due to the coherent superposition.
As argued above, the regime $|z_2 - z_1| \sim \Delta$ provides a favorable setting for observing the effect, so in total $\sigma_z \sim \Delta$, and, more generally, 
\begin{equation}
    \sigma_\Gamma / \Gamma \sim \gamma_\text{Q}^{-1}.
\end{equation}
This implies that if experimental accuracy allows for resolving $\gamma_\text{Q}^{-1}$, then the effect of the spatial spread of the wave function will have the subleading effect on the total uncertainty.

Finally, from Eq.~\eqref{beta} we can extract the shape of the emission line.
The probability that the atom emits a photon with energy $\omega_{\boldsymbol{k}}$ equals
\begin{align}
\mathcal{P}(\omega_k)= \lim_{\tau\to\infty}\sum_{\lambda=1,2}\int \mathrm{d}z|\beta_{k,\lambda}(z,\tau)|^2.
\end{align}
Assuming that $\Omega/\Gamma_0\ll1$, we arrive at the following formula for the transition line (see Appendix~\ref{appendix::emission}):
\begin{align}
&\mathcal{P}(\omega_k)=\nonumber \\
&\int \mathrm{d}z\frac{1}{2\pi}\frac{\left(1+\frac{gz}{c^2}\right)|\psi(z)|^2\Gamma_0}{\frac{1}{4}\Gamma_0^2\left(1+\frac{gz}{c^2}\right)^2+\left(\Omega\left(1+\frac{gz}{c^2}\right)-\omega_k\right)^2}.
\end{align}
Thus, $\mathcal{P}(\omega_k)$ is proportional to the height distribution $|\psi(z)|^2$ integrated against a Lorentz distribution. The transition line is gravitationally blue- or red-shifted, depending on the position of the atom, as the Lorentz distribution is shifted by $\Omega\to\Omega\left(1+\frac{gz}{c^2}\right)$. The transition line splits in two for double-peaked wave functions, with peaks at a considerable distance from each other. The difference between the shape of the emission line of a coherent superposition Eq.~\eqref{superposition} and classical mixture Eq.~\eqref{mixture} of wave packets has been plotted for specific configurations of $\psi_{\text{sup}}(z)$ and $\mathnormal{P}_{\text{cl}}(z)$ in Fig.~\ref{fig::spectral_lines}. Note that the difference in the shape of the spectrum gradually disappears when both the difference in heights $z_2-z_1$ decreases for given $\Omega/\Gamma_0$, and when the ratio $|z_2-z_1|/\Delta$ increases. This is in line with the previously described behavior of $\gamma_\text{Q}^{-1}$\,---\,the quantum time dilation effect disappears when the spatial separation of the wave packets becomes significantly larger than their spatial extent.

\section{Conclusions}
We have provided an example of a realistic situation where quantum time dilation in a gravitational field should occur. We have analyzed the spontaneous emission process of a two-level atom resting in an external gravitational field that was modeled in accordance with the equivalence principle as an accelerated frame of reference. We have shown that the spontaneous emission rate of the atom depends on its wave function in position space. In particular, this rate is influenced nontrivially by the presence of spatial coherence present in the center of the mass state of the atom. We have confirmed the result of \cite{Khandelwal2020} for a realistic clock model, providing further evidence that quantum time dilation is a universal phenomenon. Moreover, we made use of the equivalence principle to describe the effect of gravity on the clock, but the final result of our considerations is the same as would be expected from a post-Newtonian analysis. This suggests that this conclusion can be interpreted as confirmation of the equivalence principle applied to quantum systems.

Our analysis leads to a method for detecting the effect of quantum time dilation---one needs to set a decaying particle in a superposition of heights and track the dependence of the decay rate on the initial state of the particle.
Our analysis leads to a spectroscopic method for detecting the effect of quantum time dilation---one needs to set a clock (either ionic or atomic) in a superposition of heights and perform measurement either of a spontaneous decay or, more commonly, a fractional frequency shift.
In accordance with our results, the effect of coherence should be noticeable when the spread of two position wave packets becomes comparable to the distance between them. The quantum correction to classical time dilation can be (for appropriately chosen parameters of the state) of the same order of magnitude as the classical gravitational time dilation factor. Therefore, if one can detect gravitational time dilation for such distances, one should also be able to detect the quantum time dilation effect.

Usually, experimental measurements of gravitational time dilation involve comparing two clocks at different heights as achieved in tabletop experiments~\cite{Chou2010}, using flight-based clocks~\cite{Hafele1972}, or clocks separated by hundreds of meters~\cite{Takamoto2020}.
For such approaches, the next advances are already planned: satellite-based experiments that will allow researchers to improve the accuracy by orders of magnitude~\cite{Laurent2015,Tino2019}. 
Recent developments with optical lattice clocks also showed that resolving the gravitational redshift within a single sample on a sub-millimeter scale is possible~\cite{Bothwell2022,Zheng2022}.
Specifically, a change of frequency consistent with the linear gravitational field was measured along the system consisting of 100,000 strontium atoms~\cite{Bothwell2022}.
The atoms were uncorrelated to suppress corrections due to quantum coherence across the sample.

We have shown that for an optimally prepared state in the simplest spectroscopic system, the gravitational quantum time dilation effect is comparable to the gravitational redshift induced by a millimeter-sized height difference close to the Earth's surface. In the most favorable setup with two overlapping wave packets of opposite relative phase, the change in the total emission rate scales like $\frac{g \Delta}{4 c^2} \Gamma_0$, where $\Delta$ is the spatial spread of wave packets.
In the case of micrometer-scale superpositions \cite{charriere2012, zhang2016, xu2019, panda2023}, this amounts to a $10^{-23}$ change in total emission rate, or, equivalently, the same change in the fractional frequency shift. While the atomic lifetimes, and thus emission rates, are currently measured with insufficient precision to detect such a correction\,---\,they are typically determined up to tenths of a percent \cite{Rafac1994, Moehring2006, Seidlitz2011}\,---\,the correction to the fractional frequency shift is just below the precision of state-of-the-art measurements, which sensitive to gravitational time dilation millimeter scales~\cite{Hutson2019,Bothwell2022,Zheng2022,Zheng2023}. Increasing the scale of the spatial superposition of the atomic species from micrometers to millimeters would result in a correction of the order of $10^{-20}$, which is within the reach of current technology.
Given the recent experimental progress on optical clocks, the natural next step is to devise a scheme to prepare the optimal superposition state in this setting and examine how to unambiguously observe the quantum time dilation with present-day technology as the magnitude of the effect is within current experimental precision.

\begin{acknowledgments}
We would like to thank Shadi Ali Ahmad and Magdalena Zych for fruitful correspondence and useful comments.
P.~T.~G. is financed by the (Polish) National Science Center Grant 2020/36/T/ST2/00065 and supported by the Foundation for Polish Science (FNP).
The Center for Theoretical Physics of the Polish Academy of Sciences is a member of the National Laboratory of Atomic, Molecular, and Optical Physics (KL FAMO).
K.~D. is financially supported by the (Polish) National Science Center Grant 2021/41/N/ST2/01901.
A.~R.~H.~S. wishes to thank Saint Anselm College for support through a summer research grant.

\end{acknowledgments}

\bibliographystyle{quantum}
\bibliography{library}

\onecolumngrid

\appendix

\section{Evolution of the system}\label{appendix::evolution}
In this Appendix, we analyze the evolution of the atomic system in Rindler coordinates making use of the Hamiltonian
\begin{equation}
    \hat{H}=\hat{H}_{\text{atom}}+\hat{H}_{\text{field}}+\hat{H}_{\text{af}}.
\end{equation}
We need to solve the Schr{\"o}dinger equation with the above Hamiltonian and the ansatz in Eq.~\eqref{general_state}
\begin{equation}\label{schrodinger}
\begin{split}
&i\hbar\left(\int\mathrm{d}z\, \Dot{\alpha}(z,\tau)\ket{z,e,0}+\sum_{\lambda=1,2}\int\int\mathrm{d}k\,\mathrm{d}z\, \Dot{\beta}_{k,\lambda}(z,\tau)\ket{z,g,1_{k,\lambda}}\right)\\
&=\int\mathrm{d}z\left( \left(M c^{2}+\hbar\Omega\right)\left(1+\frac{gz}{c^2}\right)\alpha(z,\tau)-i\hbar\sum_{\lambda=1,2}\int\mathrm{d}k\,g_{k,\lambda}\mathrm{e}^{ik\xi}\beta_{k,\lambda}(z,\tau)\right)\ket{z,e,0}\\
&+\sum_{\lambda=1,2}\int\int\mathrm{d}k\,\mathrm{d}z\,\left(\left(M c^2\left(1+\frac{gz}{c^2}\right)+\hbar\omega_k\right)\beta_{k,\lambda}(z,\tau)-i\hbar g_{k,\lambda}\mathrm{e}^{-ik\xi}\alpha(z,\tau)\right)\ket{z,g,1_{k,\lambda}}.
\end{split}
\end{equation}
Here, the dot denotes the derivative with respect to the coordinate time $\tau$. The infinite set of equations implied by this Schr{\"o}dinger equation reads
\begin{equation}
\begin{split}
\Dot{\alpha}(z,\tau)=&-i\omega_e(z)\alpha(z,\tau)-\sum_{\lambda=1,2}\int \mathrm{d} k\,g_{k,\lambda}\mathrm{e}^{ik\xi}\beta_{k,\lambda}(z,\tau),\\
\Dot{\beta}_{k,\lambda}(z,\tau)=&-i\left(\omega_g(z)+\omega_k\right)\beta_{k,\lambda}(z,\tau)-g_{k,\lambda}\mathrm{e}^{-ik\xi}\alpha(z,\tau),
\end{split}
\end{equation}
with
\begin{equation}
\begin{split}
\omega_g(z)=\frac{Mc^2}{\hbar}\left(1+\frac{gz}{c^2}\right) \qquad \text{and} \qquad  \omega_e(z)=\left(\frac{Mc^2}{\hbar}+\Omega\right)\left(1+\frac{gz}{c^2}\right).
\end{split}
\end{equation}
The initial conditions are the following
\begin{equation}
\alpha(z,0)=\psi(z),\qquad \beta_{k,\lambda}(z,0)=0.
\end{equation}
We perform the Laplace transform and find
\begin{equation}
\begin{split}
\omega\Tilde{\alpha}(z,\omega)-\psi(z)=&-i\omega_e(z)\Tilde{\alpha}(z,\omega)-\sum_{\lambda=1,2}\int \mathrm{d} k\,g_{k,\lambda}\mathrm{e}^{ik\xi}\Tilde{\beta}_{k,\lambda}(z,\omega),\\
\omega\Tilde{\beta}_{k,\lambda}(z,\omega)=&-i\left(\omega_g(z)+\omega_k\right)\Tilde{\beta}_{k,\lambda}(z,\omega)-g_{k,\lambda}\mathrm{e}^{-ik\xi}\Tilde{\alpha}(z,\omega).
\end{split}
\end{equation}
These equations lead to the following formulas for $\Tilde{\alpha}(z,\tau)$ and $\Tilde{\beta}_{k,\lambda}$
\begin{equation}\label{laplace}
\begin{split}
\Tilde{\alpha}(z,\omega)=\frac{\psi(z)}{\mathfrak{H}(\omega)},\qquad
\Tilde{\beta}_{k,\lambda}(z,\omega)=\frac{-g_{k,\lambda}\mathrm{e}^{-ik\xi}\Tilde{\alpha}(z,\omega)}{\omega+i\left(\omega_g(z)+\omega_k\right)},
\end{split}
\end{equation}
where
\begin{equation}
\begin{split}
\mathfrak{H}(\omega)=\omega+i\omega_e(z)-\sum_{\lambda=1,2}\int \mathrm{d} k\,\frac{g_{k,\lambda}^2}{\omega+i\left(\omega_g(z)+\omega_k\right)}.
\end{split}
\end{equation}
We return to the time domain using an inverse Laplace transform with integration contour $\Upsilon$ going from negative imaginary infinity to positive imaginary infinity, closed by a large semicircle to the left of the imaginary axis
\begin{equation}
\alpha(z,\tau)=\frac{1}{2\pi i}\int_\Upsilon \mathrm{d} \omega\,\frac{\mathrm{e}^{\omega \tau}\psi(z)}{\mathfrak{H}(\omega)},
\end{equation}
and use the single pole approximation $\mathfrak{H}(\omega)=\omega-\omega_0$ with
\begin{equation}
\begin{split}
\omega_0=-i\omega_e(z)+\delta,\qquad
\delta=\sum_{\lambda=1,2}\int \mathrm{d} k\,\frac{ig_{k,\lambda}^2}{\left(\omega_k-\Omega\left(1+\frac{gz}{c^2}\right)\right)-i\varepsilon}.
\end{split}
\end{equation}
Using the Sochocki-Plemelj formula
\begin{equation}\label{sochocki}
\lim_{\epsilon\to 0^+}\frac{1}{x-i\varepsilon}=i\pi\delta(x)+\mathcal{P}\left(\frac{1}{x}\right),
\end{equation}
we transform it to
\begin{equation}
\begin{split}
\omega_0=-i\omega_e(z) -\frac{\Gamma(z)}{2},\qquad
\Gamma(z)=2\pi\sum_{\lambda=1,2}\int \mathrm{d} k\,g_{k,\lambda}^2\delta\left(\Omega\left(1+\frac{gz}{c^2}\right)-\omega_k\right).
\end{split}
\end{equation}
We note that
\begin{equation}\label{polarization_sum}
\sum_{\lambda=1,2}\hat{\boldsymbol{e}}_{\lambda,i}\hat{\boldsymbol{e}}_{\lambda,j}=\delta_{i,j}-\frac{k_ik_j}{k^2},
\end{equation}
and assume that the dipole moment of the atom is perpendicular to the direction of light propagation (direction of the gravitational field), to finally compute $\Gamma(z)$:
\begin{equation}\label{gamma}
\begin{split}
\Gamma(z)=\int \mathrm{d} \omega_k\,\frac{\omega_k\boldsymbol{d}^2}{2\hbar c\varepsilon_0}\delta\left(\Omega\left(1+\frac{gz}{c^2}\right)-\omega_k\right)=\left(1+\frac{gz}{c^2}\right)\frac{\Omega \boldsymbol{d}^2}{2\hbar c\varepsilon_0}=\left(1+\frac{gz}{c^2}\right)\Gamma_0.
\end{split}
\end{equation}
Here $\Gamma_0=\frac{\Omega \boldsymbol{d}^2}{2\hbar c\varepsilon_0}$ is the transition rate of the atom in the absence of gravity, whereas $\Gamma(z)$ is the transition rate of a particle localized at height $z$ in a gravitational field.

With Eq.~\eqref{gamma} in hand, we can calculate the amplitude $\alpha(z,\tau)$ in the single pole approximation
\begin{equation}
\begin{split}
\alpha(z,\tau)=\psi(z)\exp(\omega_0\tau)=\psi(z)\exp\left[\left(-i\omega_e(z) -\frac{\Gamma(z)}{2}\right)\tau\right].
\end{split}
\end{equation}
Now we substitute $\frac{\psi(z)}{\omega-\omega_0}$ for $\Tilde{\alpha}(z,\omega)$ in Eq.~\eqref{laplace}  for $\Tilde{\beta}_{k,\lambda}(z)$, and perform the inverse Laplace transform to obtain
\begin{equation}
\beta_{k,\lambda}(z,\tau)=\frac{g_{k,\lambda}\psi(z)\mathrm{e}^{-i\left(\omega_g(z)+k\xi\right)}}{\frac{1}{2}\Gamma(z)+i\left(\Omega\left(1+\frac{gz}{c^2}\right)-\omega_k\right)}\left[\mathrm{e}^{-\left(i\Omega\left(1+\frac{gz}{c^2}\right)+\frac{1}{2}\Gamma(z)\right)\tau}-\mathrm{e}^{i\omega_k\tau}\right].
\end{equation}

\section{Derivation of the emission rate and spectrum shape}\label{appendix::emission}
Using the results from Appendix~\ref{appendix::evolution}, one can compute the probability that the atom stays in the excited state until coordinate time $\tau$
\begin{equation}\label{survival}
\int \mathrm{d}z|\alpha(z,\tau)|^2
=\int \mathrm{d}z|\psi(z)|^2\exp\left(-\Gamma(z)\tau\right).
\end{equation}
The transition rate is defined as the time derivative of this probability
\begin{equation}\label{transition rate definition}
\begin{split}
\Gamma=-\frac{\mathrm{d}}{\mathrm{d}\tau}\int \mathrm{d}z|\psi(z)|^2\exp\left[-\Gamma(z)\tau\right]=\int \mathrm{d}z|\psi(z)|^2\Gamma(z)\exp\left[-\Gamma(z)\tau\right]\approx \int \mathrm{d}z|\psi(z)|^2\Gamma(z).
\end{split}
\end{equation}
Here in the last line, we made an assumption that the time $\tau$ is much shorter than the lifetime of the excited state in the absence of gravity $\tau\ll \left(\Gamma_0\right)^{-1}$, and we consider only the cases with $gz/c^2\ll1$ in the range of non-vanishing $\psi(z)$.
From Eq.~\eqref{transition rate definition}, we conclude that the total transition rate for a particle described by the wave function $\psi(z)$ is given by a weighted mean of transition rates at different heights. Making use of Eq.~\eqref{gamma} we can rewrite Eq.~\eqref{transition rate definition} in a form
\begin{equation}
    \Gamma \approx \int \mathrm{d}z|\psi(z)|^2\left(1+\frac{gz}{c^2}\right)\Gamma_0=\Gamma_0\left(1+\frac{g\langle z\rangle}{c^2}\right),
\end{equation}
where we used the normalization condition $\int \mathrm{d}z|\psi(z)|^2=1$, and defined $\langle z\rangle\equiv\int \mathrm{d}z|\psi(z)|^2 z$.

The variance in the transition rate is
\begin{equation}
    \sigma_\Gamma^2\equiv\int \mathrm{d}z|\psi(z)|^2 \Gamma(z)^2-\left(\int \mathrm{d}z|\psi(z)|^2 \Gamma(z)\right)^2=\Gamma_0^2\frac{g^2}{c^4}\left(\langle z^2\rangle-\langle z\rangle^2\right)\equiv\Gamma_0^2\left(\frac{g\sigma_z}{c^2}\right)^2.
\end{equation}
The normalized standard deviation of the transition rate reads
\begin{equation}
    \frac{\sigma_\Gamma}{\Gamma}\approx\frac{g\sigma_z}{c^2}.
\end{equation}

We are interested in computing the transition rate of an atom in a coherent superposition of two wave packets, $\Gamma_{\text{sup}}$, and comparing it with the transition rate of an atom in a probabilistic mixture of these wave packets, $\Gamma_{\text{cl}}$. We assume that in the first case, the initial wave function is given by
\begin{equation}\label{superposition}
\psi_{\text{sup}}(z)=\mathcal{N}\left[\cos\theta\mathrm{e}^{-\frac{(z-z_1)^2}{2\Delta^2}}+\mathrm{e}^{i\varphi}\sin\theta\mathrm{e}^{-\frac{(z-z_2)^2}{2\Delta^2}}\right],
\end{equation}
with
\begin{equation}
\mathcal{N}=\left[\sqrt{\pi}\Delta\left(1+\cos\varphi\sin2\theta\mathrm{e}^{-(z_1-z_2)^2/4\Delta^2}\right)\right]^{-1/2},
\end{equation}
whereas in the second case, the probability density reads
\begin{equation}\label{mixture}
\mathnormal{P}_{\text{cl}}(z)=\frac{1}{\sqrt{\pi}\Delta}\left[\cos^2\theta\mathrm{e}^{-\frac{(z-z_1)^2}{\Delta^2}}+\sin^2\theta\mathrm{e}^{-\frac{(z-z_2)^2}{\Delta^2}}\right].
\end{equation}
In order to compare these two transition rates we compute the ratio
\begin{equation}\label{appendix::ratio}
\begin{split}
\frac{\Gamma_{\text{sup}}-\Gamma_{\text{cl}}}{\Gamma_0}&=\int \mathrm{d}z\,\left(1+\frac{gz}{c^{2}}\right)\left(|\psi_{\text{sup}}(z)|^2-\mathnormal{P}_{\text{cl}}(z)\right) \\
&= \frac{g}{c^2}\Big(\langle z\rangle_\text{sup}-\langle z\rangle_\text{cl}\Big)\\
&=\frac{g\cos\varphi\sin4\theta(z_2-z_1)}{4c^2\left[\cos\varphi\sin2\theta+\mathrm{e}^{\frac{(z_2-z_1)^2}{4\Delta^2}}\right]}.
\end{split}
\end{equation}

The quantum superposition, Eq.~\eqref{superposition}, and classical mixture, Eq.~\eqref{mixture}, differ not only in the total transition rate but also in the shape of the associated emission spectrum. In order to check this we must compute the probability that the atom ultimately emits a photon with energy $\hbar\omega_k$
\begin{equation}\label{emission}
\mathcal{P}(\omega_k)=\lim_{\tau\to\infty}\sum_{\lambda=1,2}\int \mathrm{d}z|\beta_{k,\lambda}(z,\tau)|^2=\sum_{\lambda=1,2}\int \mathrm{d}z\frac{g_{k,\lambda}^2|\psi(z)|^2}{\frac{1}{4}\Gamma(z)^2+\left(\Omega\left(1+\frac{gz}{c^2}\right)-\omega_k\right)^2}.
\end{equation}
Substituting $g_{k,\lambda}^2=\frac{\omega_{k}}{4 \pi\hbar\varepsilon_0}(\boldsymbol{d}\cdot\hat{\boldsymbol{e}}_{\lambda})^2$, and performing the sum over the polarizations, we obtain
\begin{equation}
\mathcal{P}(\omega_k)=\int \mathrm{d}z\frac{\omega_k}{2\pi\Omega}\frac{|\psi(z)|^2\Gamma_0}{\frac{1}{4}\Gamma_0^2\left(1+\frac{gz}{c^2}\right)^2+\left(\Omega\left(1+\frac{gz}{c^2}\right)-\omega_k\right)^2}.
\end{equation}

Usually, the transition rate $\Gamma_0$ is several orders of magnitude smaller than the resonant light frequency $\Omega$. Therefore, the integrand vanishes when the value of $\omega_k$ differs significantly from $\Omega\left(1+\frac{gz}{c^2}\right)$, and we can replace $\omega_k$ in the numerator of the integrand by $\Omega\left(1+\frac{gz}{c^2}\right)$. The expression we are left with reads
\begin{equation}\label{emission1}
\mathcal{P}(\omega_k)=\int \mathrm{d}z\frac{1}{2\pi}\frac{\left(1+\frac{gz}{c^2}\right)|\psi(z)|^2\Gamma_0}{\frac{1}{4}\Gamma_0^2\left(1+\frac{gz}{c^2}\right)^2+\left(\Omega\left(1+\frac{gz}{c^2}\right)-\omega_k\right)^2},
\end{equation}
which, plotted for the quantum superposition and mixed state (see Fig.~\ref{fig::spectral_lines}), reveals the difference between these two cases.

\section{Approximated results from [Khandelwal \textit{et al.}, Quantum \textbf{4}, 309 (2020)]}\label{appendix_comparison}
In Ref.~\cite{Khandelwal2020} it was shown that a quantum clock moving with mean velocity $v_0$ and described by a superposition of two Gaussian wave packets with different mean heights, i.e.,
\begin{equation}
\ket{\psi}=\frac{1}{\sqrt{N}}\left(\sqrt{\alpha}\ket{\psi_1}+\mathrm{e}^{i\varphi}\sqrt{1-\alpha}\ket{\psi_2}\right),
\end{equation}
where $\ket{\psi_1}$ and $\ket{\psi_2}$ are Gaussian states differing only in the value of mean height $\langle\hat{z}\rangle$ (the first one is localized around $z_1$, and the second one around $z_2$), reads the average time
\begin{equation}\label{timeread}
\langle\hat{T}\rangle_{\text{sup}}(t)=\langle\hat{T}\rangle_{\text{mix}}(t)+T_{\text{coh}}(t),
\end{equation}
where $t$ is the proper time of an observer at rest at the ground level $z=0$, $\langle\hat{T}\rangle_{\text{mix}}$ is the average time read by the clock described by the classical mixture of the same two Gaussian states $\ket{\psi_1}$ and $\ket{\psi_2}$, and $\hat{T}_{\text{coh}}$ is the contribution due to coherence between these two states.
According to \cite{Khandelwal2020}, this second contribution is equal to
\begin{equation}\label{full_correction}
\begin{split}
 T_{\text{coh}}(t)=\frac{N-1}{2N}\left[\left(\frac{z_2-z_1}{2\sigma_z}\right)^2\frac{\sigma_v^2}{c^2}-\frac{g(z_2-z_1)}{c^2}(1-2\alpha)-\frac{2}{\hbar}\frac{\sigma_v^2}{c^2}(z_2-z_1)(\boldsymbol{\bar{p}}-mgt)\tan\varphi\right]t.
\end{split}
\end{equation}
Here $\sigma_z$ and $\sigma_v$ are the standard deviations in position and velocity of $\ket{\psi_1}$ and $\ket{\psi_2}$, respectively, $\boldsymbol{\bar{p}}$ is the mean momentum, and the normalization factor $N$ is equal to
\begin{equation}
N=1+2\cos\varphi\sqrt{\alpha(1-\alpha)}\mathrm{e}^{-\left(\frac{z_2-z_1}{2\sigma_z}\right)^2},
\end{equation}
(notice that with such a normalization factor, the state is normalized to $\braket{\psi}{\psi}=\sqrt{\pi}\sigma_z$).
Let us estimate the order of magnitude of individual terms from Eq.~\eqref{full_correction} for parameters used to plot Fig.~\ref{fig::spectral_lines}. For instance, if we take the difference of heights $z_2-z_1\sim 10^{-18}c^2/g$, the height dispersion $\sigma_z\sim 10^{-18}c^2/g$, the mass of the atom $m\sim 1\text{u}\sim 10^{-27}\text{kg}$, use the fact that $\sigma_z\sigma_p\sim\hbar$, and recall that $\sigma_p=m\sigma_v$, we get
\begin{equation}
   \frac{\sigma_v^2}{c^2}\sim\frac{\hbar^2}{m^2c^2\sigma_z^2}\sim 10^{-26}.
\end{equation}
The first term in the bracket in Eq.~\eqref{full_correction} is then of the order $\sim 10^{-26}$, whereas the second one is $\sim 10^{-18}$. To estimate the third term we recall that we consider a resting atom, i.e., $\boldsymbol{\bar{p}}=0$, and we compute the transition rate at times much smaller than a spontaneous emission lifetime of the excited state in the absence of gravitational field $t\ll\Gamma_0^{-1}$. Typically we have $\Gamma_0^{-1}\sim 10^{-8}s$, which means that the factor multiplying $\tan\varphi$ in the third term cannot be greater than $\sim 10^{-26}$. We should stress that the tangent function appearing in the last term does not lead to any infinities for $\varphi\to\pi/2$, because for such $\varphi$ the factor multiplying the whole bracket vanishes, and the overall result is finite and relatively small (compared to the value at $\varphi=0$ or $\varphi=\pi$). Therefore we can neglect both the first and the third term to obtain
\begin{equation}\label{correction}
\begin{split}
 T_{\text{coh}}(t)=\frac{N-1}{2N}\frac{g(z_2-z_1)}{c^2}(2\alpha-1)t=\frac{g\cos\varphi\sqrt{\alpha(1-\alpha)}(2\alpha-1)(z_2-z_1)}{c^2\left(2\cos\varphi\sqrt{\alpha(1-\alpha)}+\mathrm{e}^{\left(\frac{z_2-z_1}{2\sigma_z}\right)^2}\right)}t\equiv\gamma^{-1}_\text{Q}t.
\end{split}
\end{equation}
The omission of the terms proportional to $\sigma_v^2/c^2$ in this paper can be traced back to the fact that we omitted all the kinetic terms in the atomic Hamiltonian in Eq.~\eqref{atomic_hamiltonian}, so that we completely neglect any motion of the atom, and concentrate on the purely gravitational effect. The estimation presented above can be treated as a justification for this omission of the considered range of parameters.

Let us rewrite Eq.~\eqref{correction} in a slightly different notation, as used in the present paper. We substitute $\alpha\to\cos^2\theta$, and $\sigma_z\to\Delta$, to get
\begin{equation}
\begin{split}
 \gamma^{-1}_\text{Q}&=\frac{g\sqrt{\cos^2\theta(1-\cos^2\theta)}(2\cos^2\theta-1)(z_2-z_1)}{c^2\left(\mathrm{e}^{\left(\frac{z_2-z_1}{2\Delta}\right)^2}+2\cos\varphi\sqrt{\cos^2\theta(1-\cos^2\theta)}\right)}
=\frac{g}{4c^2} \frac{\cos\varphi\sin4\theta(z_2-z_1)}{\cos\varphi\sin2\theta+\mathrm{e}^{\frac{(z_2-z_1)^2}{4\Delta^2}}} .
\end{split}
\end{equation}
This is the same expression that appears in Eq.~\eqref{ratio}.

\end{document}